\documentclass{optica-article}

\journal{opticajournal} 

\articletype{Research Article}

\usepackage{lineno}

\begin{document}

\title{Efficient single-photon directional transfer between waveguides via two giant atoms}

\author{Daqiang Bao,\authormark{1} Zhirong Lin,\authormark{1,2,*}}

\address{\authormark{1}State Key Laboratory of Materials for Integrated Circuits, Shanghai Institute of Microsystem and Information Technology, Chinese Academy of Sciences, 200050 Shanghai, China\\
\authormark{2}University of Chinese Academy of Science, 100049 Beijing, China}
\email{\authormark{*}zrlin@mail.sim.ac.cn} 


\begin{abstract*} 
We investigate the single-photon transport properties in a double-waveguide quantum electrodynamic system. We force the energy degeneracy of the collective states by adjusting the direct coupling strength between the two giant atoms. Our results indicate that resonant photons can be completely transferred between the two waveguides owing to the scattering interference of eigenstates, which also results in the directional propagation of resonant photons in the output waveguide. Perfect transfer occurs when the two scattering states degenerate in the energy and decay rates. We further propose a simple scheme to realize the efficient photon transfer with directional control. This study has potential applications in quantum networks and integrated photonic circuits.
\end{abstract*}

\section{Introduction}
Quantum information transfer between nodes is the foundation of extensible quantum processor networks\cite{kimble,tra1}. Single photons are excellent carriers of quantum information owing to their high-speed transmission and low-noise properties\cite{pho}. Single-photon routers are essential for quantum networks. The key to a quantum router is the control of single-photon transport between channels. Given that waveguides serve as quantum channels for photons, systems based on waveguide quantum electrodynamics (QED) provide ideal platforms for investigating photon transport properties\cite{rmp1}. In waveguide-QED systems, quantum emitters are strongly coupled to the continuum of propagating photonic modes\cite{w1}. Strong interactions not only enable many quantum optical phenomena to be observed in experiments, such as resonance fluorescence\cite{mollow,mollow1}, super- and sub-radiance\cite{sp1,sp2,sp3,sp4,sp5}, and Lamb shifts\cite{lamb,lamb1}, but also influence photon transport in waveguides\cite{fan1}. Numerous studies have explored quantum routing and photon transport in a single waveguide by controlling multi-level atoms\cite{r1,r2,r3} and multi-atoms\cite{w1,mollow1,r6,r81,r82,r83,r84,Rev2}. 

The multi-waveguide system offers additional photon information channels, facilitating the integration of photon-based quantum technologies\cite{io}.
However, the transfer efficiency in the conventional bidirectional waveguides may be low due to losses associated with undesired directional components. To overcome this problem, many proposals have recently been proposed for double-waveguide systems, such as chiral coupling\cite{r3,r14,r15} or interference between channels\cite{r9,r10,r11,r12,r13,Rev1}. Previous research has suggested that complete photon transfer between continua can be achieved by creating resonant states with different symmetries, and forcing an accidental degeneracy between them\cite{fand2}. The recent studies also showed directional microwave photon emission in an energy degenerate waveguide system\cite{on1,Rev3}.

Inspired by these advancements, this study explores the single-photon transfer properties between two waveguides. We propose a system where the collective states of two giant atoms are deliberately engineered to achieve energy degeneration. We analytically calculated single-photon scattering amplitudes. The resulting scattering spectra demonstrate that complete photon transfer can be achieved with the directionality of the transferred photons being contingent upon the separation distance between the two giant atoms. Furthermore, a channel-drop filter effect becomes evident when the giant atoms are positioned a quarter wavelength apart.

\section{Model and theory}\label{2}
\begin{figure}[htb]
    \centering\includegraphics[width=7cm]{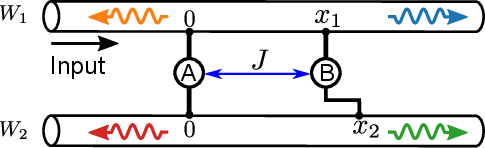}
    \caption{\label{fig1}Configuration of two giant atom-waveguide system. The giant atom labeled A couples to the $W_1$ and $W_2$ at the origin of coordinates. The giant atom labeled B couples to  $W_1$ and $W_2$ at points $x_1$ and $x_2$, respectively. $J$ is the external coupling strength between giant atoms A and B.}
\end{figure}
We consider a waveguide-QED system with two giant artificial atoms and two waveguides. As shown in Fig.\ref{fig1}, two giant atoms are identical and assumed as a two-level system with the ground state $|g\rangle$ and excited state $|e\rangle$. The giant atom A is placed at the coordinate origin of waveguide 1 ($W_1$) and waveguide 2 ($W_2$). The giant atom B is coupled to the waveguides at $x_1$ in $W_1$ and $x_2$ in $W_2$. The Hamiltonian of the waveguide QED system reads ($\hbar=1$):
\begin{align}
    H&=iv_g\sum_{i=1,2}\int{dx}\left(a_{Li}^\dag(x)\frac{\partial}{\partial x}a_{Li}(x)-a_{Ri}^\dag(x)\frac{\partial}{\partial x}a_{Ri}(x)\right)\nonumber\\	
	&+V\sum_{i=1,2}\int dx\delta(x)\left[S_a^\dag\left(a_{Ri}(x)+a_{Li}(x)\right)+H.c.\right]\nonumber\\
	&+V\sum_{i=1,2}\int dx\delta(x-x_i)\left[S_b^\dag\left(a_{Ri}(x)+a_{Li}(x)\right)+H.c.\right]\nonumber\\
	&+(\omega-i\kappa/2)( S_a^\dag S_a^-+S_b^\dag S_b^-)+J(S_a^\dag S_b^-+S_b^\dag S_a^-),\label{H}
\end{align}
where $v_g$ denotes the group velocity of the waveguide field. $a_{Li}^\dag(x)[a_{Li}(x)]$ and $a_{Ri}^\dag(x)[a_{Ri}(x)]$ describe the creation (annihilation) of left- and right-propagating photons at position $x$ in waveguide $i$. $\omega$ and $\kappa$ denote the transition frequency of giant atom and its dissipation rate into non-waveguide mode, respectively\cite{r83,r10,r13,r14}. In the strong coupling waveguide-QED system, $\kappa$ is typically very small. $J$ is the direct coupling strength between giant atoms. $S_a^\dag$ ($S_b^\dag$) and  $S_a^-$ ($S_b^-$) are the raising and lowering operators of giant atom $A$ ($B$). $V$ represents the coupling strength between the waveguide and the giant atoms. The coupling strengths are assumed to be equal. $\delta(x)$ denotes the Dirac delta function. In single-excitation subspace, the scattering eigenstate can be written as:
\begin{align}
	|\psi\rangle &=\sum_{i=1,2}\int{dx}\left(\phi_{Ri}(x)a_{Ri}^\dag(x)+\phi_{Li}(x)a_{Li}^\dag(x)\right)|0,g,g\rangle\nonumber\\
	&+(C_aS_a^\dag+C_bS_b^\dag)|0,g,g\rangle,\label{wavefunction}
\end{align}
where $\phi_{Ri}(x)$ and $\phi_{Li}(x)$ are the scattering amplitudes of the right- and left-propagating modes in waveguide $i$. $|0,g,g\rangle$ is a vacuum state in which two giant atoms are in their ground states. $C_a$ ($C_b$)
is the probability amplitude of state $|0,e,g\rangle$ ($|0,g,e\rangle$). Assuming that a single photon is incident from the left side of W1, the wave function in Eq.(\ref{wavefunction}) can be expressed as:
\begin{equation}\label{ewave}
    \begin{split}
        \phi_{R1}&=e^{ikx}[H(-x)+t_1H(x)H(x_1-x)+t_{R1}H(x-x_1)],\\
        \phi_{L1}&=e^{-ikx}[r_{L1}H(-x)+r_1H(x)H(x_1-x)],\\
        \phi_{R2}&=e^{ikx}[t_2H(x)H(x_2-x)+t_{R2}H(x-x_2)],\\
        \phi_{L2}&=e^{-ikx}[r_{L2}H(-x)+r_2H(x)H(x_2-x)].
    \end{split}
\end{equation}
Here, $H(x)$ is the Heaviside step function with $H(0)=1/2$. The wave vector $k$ ($k>0$) of the photon satisfies $E=v_gk$.  Substituting Eq.(\ref{ewave}) into the time-independent Schr\"{o}dinger equation $H|\psi\rangle=E|\psi\rangle$, we obtain the probability amplitudes:
\begin{align}
	t_{R1}&=\frac{(\Delta+i\gamma+i\kappa/2)^2-(J_C+i\gamma\cos\theta)^2+\gamma^2\sin^2\theta}{{\Delta^\prime}^2-J_C^2},\label{t0}\\
	r_{L1}&=\frac{2\gamma e^{i\theta}(J_C+\Delta^\prime\cos\theta)}{i({\Delta^\prime}^2-J_C^2)},\label{r0}\\
	t_{R2}&=\frac{\gamma[J_C(e^{i\theta}+e^{-i\phi})+\Delta^\prime(e^{i(\theta-\phi)}+1)]}{i({\Delta^\prime}^2-J_C^2)},\label{tf0}\\
	r_{L2}&=\frac{\gamma[J_C(e^{i\theta}+e^{i\phi})+\Delta^\prime(e^{i(\theta+\phi)}+1)]}{i({\Delta^\prime}^2-J_C^2)},\label{tb0}
\end{align}
where $\Delta=E-\omega$, $\theta=kx_1$, $\phi=kx_2$, $\gamma=V^2/v_g$, $J_C=J-i\gamma(e^{i\theta}+e^{i\phi})$ and
$\Delta^\prime=\Delta+2i\gamma+i\kappa/2$. 

\section{Results and analysis}\label{3}
\begin{figure}[htb]
    \centering\includegraphics{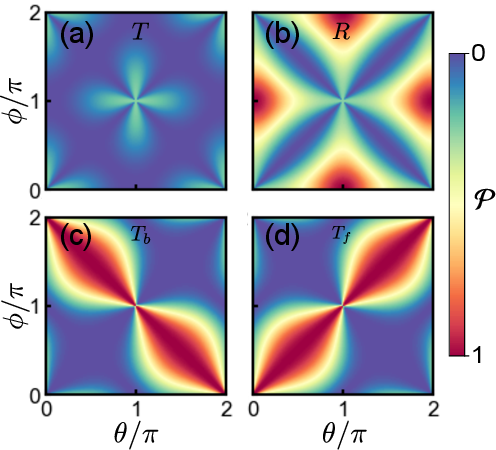}
    \caption{\label{fig2}  The scattering probability $\mathcal{P}$ ($\mathcal{P}=T, R, T_f$ and $T_b$) as a function of normalized phase shift $\theta/\pi$ and $\phi/\pi$. In (a-d), the parameters are $J=-\gamma(\sin\theta+\sin\phi)$, $\Delta=0$ and $\kappa=0$.}
\end{figure}
For $W_1$, the transmission and reflection probabilities are defined as $T=|t_{R1}|^2$ and $R=|r_{L1}|^2$. Because the photon can also be transferred to $W_2$, we define the backward and forward transfer probabilities as $T_b=|r_{L2}|^2$ and $T_f=|t_{R2}|^2$. The dissipation rate $\kappa$ is temporarily ignored in order to demonstrate photon transfer more clearly.  When $\Delta=0$, to realize complete transfer from $W_1$ to $W_2$, the complex numerator of Eq.(\ref{t0}) and Eq.(\ref{r0}) should be forced to zero. A necessary condition can be derived, i.e. the real part of the numerator is zero, which is given by the equation: $J+\gamma(\sin\theta+\sin\phi)=0$. Under these conditions, we plot the scattering probabilities as functions of $\theta/\pi$ and $\phi/\pi$ in Fig.\ref{fig2}.

The transmission coefficient $T$ [Fig.\ref{fig2}(a)] of a resonant photon is typically suppressed by scattering, which means that the photon is scattered in other directions. When $\theta=n\pi$ ($n$ is an integer) and $|\phi-\theta|=\pi$, the reflection coefficient $R$ [Fig.\ref{fig2}(b)] shows the maximum value. The minimum values of the reflection coefficient appear at $\phi=2\pi-\theta$ and $\phi=\theta$. Significantly, as demonstrated in Fig.\ref{fig2}(c) and (d), the conditions that yield the minimum reflection probability also coincide with the maximum transfer efficiency of the signal from $W_1$ to $W_2$ in a specific  direction. To clearly explore the scattering effect, we reorganize Eq.(\ref{t0}-\ref{tb0}) under the condition of $\kappa=0$, and have:
\begin{align}
	t_{R1}&=1+\frac{\gamma  (1-\cos\theta)}{  i(\Delta-\lambda_-)}+\frac{\gamma  (1+\cos\theta)}{i(\Delta-\lambda_+)},\label{t}\\
	r_{L1}&=\frac{\gamma(1-e^{i \theta})^2}{2i(\Delta-\lambda_-)}+\frac{\gamma(1+e^{i \theta})^2}{2i(\Delta-\lambda_+)},\label{r}\\
	t_{R2}&=\frac{\gamma(1-e^{i\theta})(1-e^{-i\phi})}{2i(\Delta-\lambda_-)}+\frac{\gamma(1+e^{i\theta})(1+e^{-i\phi})}{2i(\Delta-\lambda_+)},\label{tf}\\
	r_{L2}&=\frac{\gamma(1-e^{i\theta})(1-e^{i\phi})}{2i(\Delta-\lambda_-)}+\frac{\gamma(1+e^{i\theta})(1+e^{i\phi})}{2i(\Delta-\lambda_+)},\label{tb}
\end{align}
where $\lambda_\pm=\pm J_\Sigma- i(\Gamma_{1\pm}+\Gamma_{2\pm})$, the real and imaginary parts of which correspond to the energy and decay rate of the dressed state $|+\rangle$ and $|-\rangle$, respectively. Here, the dressed states of two giant atoms are defined as $|\pm\rangle=(|eg\rangle\pm|ge\rangle)/\sqrt{2}$.
The collective decay rate is expressed as $\Gamma_\pm=\Gamma_{1\pm}+\Gamma_{2\pm}$, where $\Gamma_{1\pm}=\gamma(1\pm\cos\theta)$ and $\Gamma_{2\pm}=\gamma(1\pm\cos\phi)$. $\Gamma_{i\pm}$ ($i=1,2$) represents the decay rate of $|\pm\rangle$ to $W_i$, which can characterize radiance nature of state $|\pm\rangle$, such as super-radiance or sub-radiance. $J_\Sigma=J+J_1+J_2$ is the effective coupling strength, and $J_1=\gamma\sin\theta$ ($J_2= \gamma\sin\phi$) is the coherent exchange interaction through virtual photons in $W_1$ ($W_2$)\cite{w1,mollow1}. 

\begin{figure}[htb]
    \centering\includegraphics{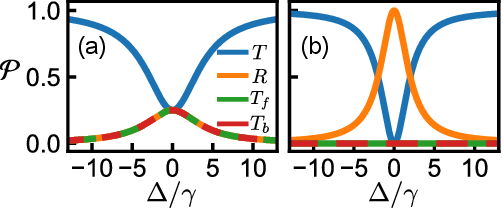}
    \caption{\label{fig3} The scattering probability as a function of normalized detuning $\Delta/\gamma$ for the different phases:(a) $\theta=\phi=\pi$ and (b) $\theta=\pi,\phi=2\pi$. The scattering probabilities at different ports are shown by different colors. $T$ is represented by the blue curves, $R$ is represented by the orange curves, $T_f$ is denoted by the green curves, and $T_b$ is shown by the red curves.}
\end{figure}
The superposition of the two eigenstate scattering amplitudes, as depicted in Eq.(\ref{t}-\ref{tb}), suggests that adjusting the phase can induce interference phenomena. On the other hand, we can always force the energy degeneracy of two scattering states by eliminating the coherent interaction, that is, $J=-(J_1+J_2)$ or $J_\Sigma=0$. This implies that interference occurs solely at resonance, where $\Delta=0$, aligning with the previously identified necessary condition for complete transfer. In subsequent discussion, we concentrate on scenarios where $J_\Sigma=0$ and $\theta,\phi\in[0,2\pi]$. 

In Fig.\ref{fig3}, we plot the scattering probability $\mathcal{P}$ as a function of the detuning. When $\theta=\phi=\pi$, $\Gamma_{1+}=\Gamma_{2+}=0$, this indicates that the state $|+\rangle$ is uncoupled from waveguide. Consequently, the only scattering path originates from the state $|-\rangle$. Therefore, interference does not occur and $r_{L1}=r_{R2}=r_{L2}$. The undirectional scattering leads to $T=R=T_b=T_f=1/4$ at $\Delta=0$. The spectra show a Lorentzian line shape with width $2\Gamma_{1-}$. When $\theta=\pi$ and $\phi=2\pi$, $W_1$ only couples to $|-\rangle$, while $W_2$ can only couple to $|+\rangle$. Therefore, signal in $W_1$ cannot be transferred to $W_2$. In Fig.\ref{fig3}(b), the orange line ($R$) shows a Lorentzian line shape with a width of $2\Gamma_{1-}$. The probabilities of scattering to $W_2$ (shown by green and red lines) are zero over the entire frequency range.

\begin{figure}[htb]
    \centering\includegraphics{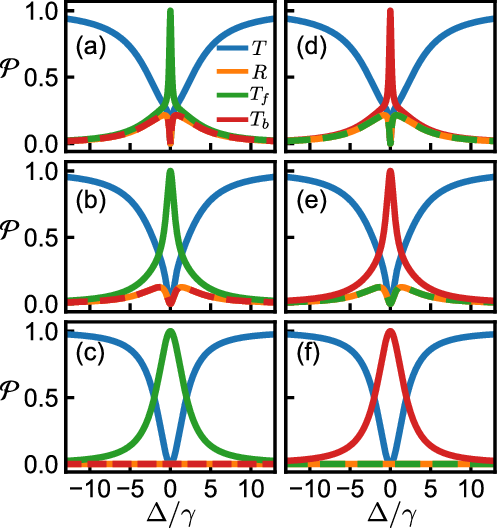}
    \caption{\label{fig4}The scattering probability as a function of normalized detuning $\Delta/\gamma$ for the different phases: (a) $\theta=\phi=\pi/8$, (b) $\theta=\phi=\pi/4$, (c) $\theta=\phi=\pi/2$, (d) $\theta=2\pi-\phi=\pi/8$, (e) $\theta=2\pi-\phi=\pi/4$ and (f) $\theta=2\pi-\phi=\pi/2$. The probabilities represented by different colored lines are same as in Fig.\ref{fig3}.}
\end{figure}
According to the above analysis, a complete transfer requires two different scattering processes to cancel the transmission and reflection in $W_1$ through destructive interference. In addition, the phase $\phi$ in $W_2$ should match the phase $\theta$ in $W_1$. When $\phi=\theta$ and $\theta\neq\pi$, the Eq.(\ref{t}) and Eq.(\ref{tb}) can be rewritten as:
\begin{align}
	r_{L1}&=r_{L2}=\frac{e^{i\theta}\Gamma_{1+}}{i\Delta-2\Gamma_{1+}}-\frac{e^{i\theta}\Gamma_{1-}}{i\Delta-2\Gamma_{1-}},\label{r2}\\
	t_{R2}&=\frac{\Gamma_{1-}}{i\Delta-2\Gamma_{1-}}+\frac{\Gamma_{1+}}{i\Delta-2\Gamma_{1+}},\label{tf2}\\
	t_{R1}&=1+t_{R2}.\label{t2}
\end{align}
The transmission and reflection are canceled at the resonance, and the photon is completely transferred to $W_2$ in forward direction. 
In Fig.\ref{fig4}(a) and (b), we plot the probabilities for $\phi=\theta=\pi/8$ and $\phi=\theta=\pi/4$. At resonance, the transmission coefficient $T$ (blue line), reflection coefficient $R$ (orange line), and backward coefficient $T_b$ (red line) exhibit a significant destructive interference dip. In contrast, the forward probability $T_f$ (green line) shows a constructive interference peak. Furthermore, the green line in Fig.\ref{fig4} (a) shows a very sharper peak than that shown in Fig.\ref{fig4} (b). We have demonstrated that the interference arises from the superposition of eigenstate ($|\pm\rangle$) scattering amplitudes. Therefore, interference mainly occurs in the overlap of scattering probabilities. This region is defined by whichever is smaller: $\Gamma_{1+}$ or $\Gamma_{1-}$. When $\theta$ is close to $0$, the decay rate $\Gamma_{1-}$ tends to $0$, leading to an extremely narrow overlapping area. The spectral line shape shows a narrower peak superimposed on a broad Lorentzian line, indicating the different radiant properties of $|+\rangle$ and $|-\rangle$. This also suggests that perfect interference occurs when scattering probabilities overlap completely, i.e., $\Gamma_{1-}=\Gamma_{1+}$. 

As shown in Eq.(\ref{r2}), if $\Gamma_{1-}=\Gamma_{1+}$, $r_{L1}$ and $r_{L2}$ are always equal to $0$, regardless of the frequency. This means that the two eigenstates are completely degenerate in terms of both energy and decay rate, i.e., $\lambda_+=\lambda_-$. This is a condition for achieving a channel drop filter\cite{fand2}. As the scattering probabilities overlap exactly, destructive interference results in complete cancellation of the reflected and backward signals. As shown in Fig.\ref{fig4}(c), the reflected wave (orange line) in $W_1$ and the backward signal in $W_2$ vanish over the entire frequency range. The transmission drops to $0$ at resonance, and the forward transferred signal exhibits a Lorentzian line shape with a width $4\gamma$. The resonant photons in $W_1$ are perfectly transferred to $W_2$. 
In Fig.\ref{fig4}(d-f), we plot the probabilities for $\theta=\pi/8$, $\theta=\pi/4$ and $\theta=\pi/2$, and $\phi=2\pi-\theta$. The red line indicates that the resonant photon transferred from $W_1$ propagates backward in $W_2$. The interpretation of this scenario parallels the analysis presented above. However, it is important to note that direct coupling is not necessary in this instance. Coherent coupling is eliminated due to $J_1=-J_2$. The effective strength of waveguide-mediated coupling is always zero under this condition. When $\theta=\pi/2$ and $\phi=3\pi/2$, we have $r_{L1}=t_{R2}$ and $t_{R1}=1+r_{L2}$. This condition can realize the backward transport of photons transferred to $W_2$. These results suggest that efficient photon transfer can be achieved and the direction can be controlled.

\begin{figure}[htb]
    \centering\includegraphics{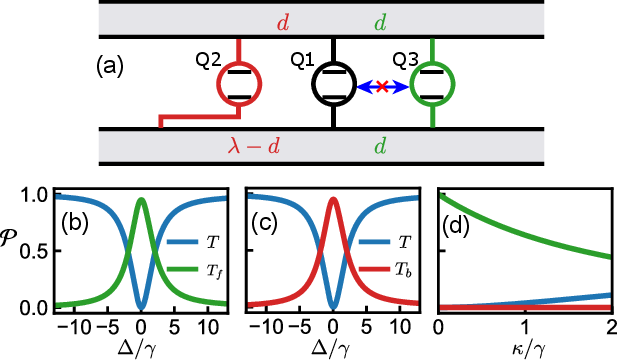}
    \caption{\label{fig5} (a) Schematic of directional photon transfer. Two waveguides are coupled via three qubits, a fixed frequency qubit (Q1) and two tunable qubits (Q2,Q3). The scattering probability is shown at detuning (b) $\delta_2=50\gamma$, $\delta_3=0$ and (c) $\delta_2=0$, $\delta_3=50\gamma$. (d) The scattering probability as a function of dissipation rate $\kappa/\gamma$ at $\delta_2=50\gamma$. $\delta_i~(i=2,3)$ is the detuning between Q$i$ and Q1. }
\end{figure}

Herein, we present a controllable directional transfer scheme within a solid system utilizing superconducting qubits. The frequency of the transmon qubit can typically be adjusted by altering the external flux through the SQUID loop. As illustrated in Fig.\ref{fig5}(a), the Q1 is a fixed-frequency qubit, whereas Q2 and Q3 are frequency tunable qubits. Moreover, a tunable coupler is required to cancel the interaction between Q2 and Q3. The actual physical parameters of the transmon qubits and coupler can be referred to the recent experimental work \cite{on1}, which demonstrates coupling control and directional photon emission with low dissipation rates. In the implementation of the protocol, one can always adjust either Q2 or Q3 to deviate from the resonant frequency of Q1, while tuning the other to be in resonance with Q1. When Q2 is significantly detuned from the resonant frequency, with Q1 and Q3 in resonance—thereby decoupling Q2 from Q1 and Q3—a forward drop filter is achieved. The scheme is also scalable by introducing more qubit interactions\cite{fand1}. The numerical results are illustrated in Fig.\ref{fig5}(b-d) for $d=\lambda/4$.
Similarly, a backward transfer can also be achieved by controlling the frequency as shown in Fig.\ref{fig5}(c). To approach the real physical system, we have considered the decay rate $\kappa=0.1\gamma$ into the non-waveguide mode in Fig.\ref{fig5}(b-c). Furthermore, the impact of dissipation is presented in Fig.\ref{fig5}(d). This result is consistent with the usual conclusion that transfer efficiency decreases as the dissipation rate increases. When $\Delta=0$, $\theta=\phi=\pi/2$ and $J_\Sigma=0$, Eq.(\ref{t0}-\ref{tb0}) indicate that $T=\kappa^2/(4\gamma+\kappa)^2$,
 $T_f=16\gamma^2/(4\gamma+\kappa)^2$ and $R=T_b=0$, which implies that this device still has good directivity.

\section{Conclusion}\label{4}
In summary, we have demonstrated the efficient directional single-photon transfer from one waveguide to another. We force the eigenenergy degeneracy by adjusting an additional coupler to eliminate waveguide-induced coherent interactions. We demonstrate the cancellation of the reflected signal due to destructive interference at resonance, ensuring that the resonant photon is completely transferred to the second waveguide. The directional propagation in the output waveguide is evidenced by analyzing the scattering interference patterns of the eigenstates. Perfect transfer is achieved when the eigenstates degenerate at both the energy and decay rate. We further propose a simple scheme based on the superconducting qubits to achieve a controlled directional photon transfer. Our study has potential applications in quantum networks and integrated photonic circuits.

\begin{backmatter}
\bmsection{Funding}
Shanghai Technology Innovation Action Plan Integrated Circuit Technology Support Program (22DZ1100200); National Key Research and Development Program of China (2023YFB4404904). 
\bmsection{Disclosures} 
The authors declare no conflicts of interest.
\bmsection{Data Availability Statement} 
The simulation data and calculations in this study can be obtained by contacting authors.
\end{backmatter}
\bibliography{reference}
\end{document}